\documentclass[aps,prb,twocolumn,showpacs,amsmath,amssymb,groupedaddress,floatfix,longbibliography]{revtex4-1}
\usepackage{graphicx,color}
\usepackage{amsmath,amssymb,bm}
\usepackage{epstopdf}
\usepackage{braket}

\setcitestyle{square,numbers}

\usepackage[plainpages=false,pdfpagelabels,colorlinks=true,linkcolor=red,urlcolor=blue,citecolor=blue,pdftitle={Title},pdfauthor={},pdfdisplaydoctitle=true,pdfduplex=DuplexFlipLongEdge]{hyperref}

\begin{document}
\title{
Density-matrix renormalization group study of the linear conductance 
in quantum wires coupled to interacting leads or phonons}

\author{Jan-Moritz Bischoff}
\email[E-mail: ]{jan.bischoff@itp.uni-hannover.de}
\author{Eric Jeckelmann}
\email[E-mail: ]{eric.jeckelmann@itp.uni-hannover.de}
\affiliation{Leibniz Universit\"{a}t Hannover, Institut f\"{u}r Theoretische Physik, Appelstra\ss e 2, D-30167 
Hannover, Germany}

\date{Draft of \today}

\begin{abstract}
In a previous paper [J.-M. Bischoff and E. Jeckelmann, Phys. Rev. B \textbf{96}, 195111 (2017)] 
we introduced a density-matrix renormalization group method for calculating the linear conductance of 
one-dimensional correlated quantum systems
and demonstrated it on homogeneous spinless fermion chains with impurities.
Here we present extensions of this method to inhomogeneous systems, models with phonons,
and the spin conductance of electronic models. The method is applied to a spinless fermion wire-lead model,
the homogeneous spinless Holstein model, and the Hubbard model. Its capabilities are demonstrated 
by comparison with the predictions of Luttinger liquid theory combined with
Bethe Ansatz solutions and other numerical methods. 
We find a complex behavior for quantum wires coupled to interacting leads 
when the sign of the interaction (repulsive/attractive) differs in wire and leads.
The renormalization of the conductance given by the Luttinger parameter in purely fermionic systems
is shown to remain valid in the Luttinger liquid phase of the Holstein model with phononic degrees of freedom.

\end{abstract}


\maketitle

\section{Introduction}

Confining electrons to one dimension results in various surprising properties because of the fermionic nature of the 
particles~\cite{baeriswyl04,giamarchi03,kawa07}. 
In practice, one-dimensional electron systems can be realized in semiconductor wires~\cite{tarucha95a}, 
atomic wires deposited on a substrate~\cite{blumenstein11a,ohtsubo15a}, carbon nanotubes~\cite{bockrath99a},
or atomic and molecular wire junctions~\cite{Agra02a,Agra02b,Ness02,Hiha12}. 
In contrast to the behavior of a three dimensional metal, which is well described by the Fermi liquid paradigm, 
theory predicts that one-dimensional conductors should exhibit the behavior of a Luttinger liquid~\cite{giamarchi03}.
In particular, the transport properties have been studied and controversially discussed
for more than three decades~\cite{giamarchi03,kawa07}.

In a previous paper~\cite{Bischoff2017}
we introduced a density-matrix renormalization group method (DMRG) for calculating the linear conductance of 
one-dimensional correlated quantum systems with short-range interactions at zero-temperature based on the Kubo formalism~\cite{kubo57a,Bohr2006}.
By taking advantage of the area law for the entanglement entropy~\cite{Eisert2010} 
the DMRG method provides us with the most efficient method for calculating the properties of 
these systems~\cite{whit92b,whit93a,scho11,jeck08a}. 
Our method combines DMRG with a finite-size scaling of dynamical correlation functions to 
compute the conductance in the thermodynamic limit.
The method was demonstrated on the homogeneous spinless fermion chains with impurities~\cite{Bischoff2017},
for which well-established results were available~\cite{giamarchi03,apel82a,kane92a,Kane1992}.

In this paper we develop the method further to treat a variety of complexer models. 
We present extensions to wire-lead systems with different interactions in the wire and leads,
to electron-phonon models, and to spinfull fermions (electrons). 
These extensions are necessary steps toward future studies of more realistic models that
are able to describe quantitatively
the experimental realizations of one-dimensional electronic 
conductors~\cite{tarucha95a,blumenstein11a,ohtsubo15a,bockrath99a,Agra02a,Agra02b,Ness02,Hiha12}. 
Here, we apply our method to a spinless fermion wire-lead model, the homogeneous spinless Holstein model, and the Hubbard model
to demonstrate its current capabilities and limitations using the predictions of Luttinger liquid theory combined with
Bethe Ansatz solutions and numerical results found in the literature.

The method is summarized in the next section. The extension and results for inhomogeneous systems such as 
the spinless fermion wire-lead system are presented in Sec.~\ref{sec:wire-lead}.
The extension to systems with phonon degrees of freedom and the results for the spinless Holstein model
are shown in Sec.~\ref{sec:holstein}.
Section~\ref{sec:hubbard} describes the generalization to spinfull fermion (electron) model and to the calculation of the spin 
conductance as well as our results for the spin conductance of the Hubbard model.
Finally, Sec.~\ref{sec:conclusion} contains our conclusion and outlook.

\section{Method}

\begin{figure}
\includegraphics[width=.99\columnwidth]{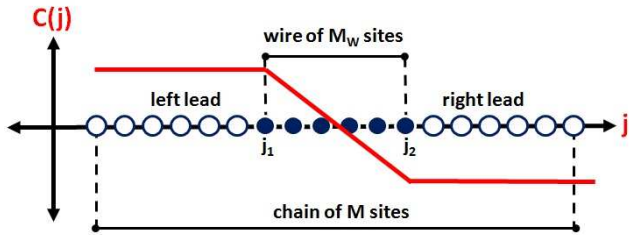}
\caption{
Schematic of the wire-lead layout for calculating the conductance. 
The chain of $M$ sites is made of left and right leads and a central wire of $M_{\text{W}}$ sites.
These three different sections are defined by the profile $C(j)$ of the external perturbation (e.g., applied bias voltage).
Additionally, wire and leads may include different sites, degrees of freedom, or interactions represented
by the solid and empty circles, respectively.
}
\label{fig:setup}
\end{figure}

In this section we summarize our DMRG method for calculating the conductance of one-dimensional correlated quantum systems.
Full details can be found in our previous paper~\cite{Bischoff2017}.
We consider a one-dimensional lattice of $M$ sites. It consists of left and right segments (called leads)
with approximately $(M-M_{\text{W}})/2$ sites
each and a central segment (called wire) of $M_{\text{W}}$ sites. A current can be generated by applying a potential bias. We assume that the potential 
remains constant in the leads and drop linearly in the wire, i.e. it has the shape
\begin{equation}
C(j)= 
  \begin{cases} 
      \hfill \hphantom{-} \frac{1}{2}     \hfill & \text{ for $j \leq j_1$} \\
      \hfill -\frac{j-j_{1}}{j_{2}-j_{1}}+\frac{1}{2}   \hfill & \text{ for $j_1 \leq j \leq j_2$} \\
      \hfill -\frac{1}{2}  \hfill & \text{ for $j \geq j_2$} 
  \end{cases}
\label{eq:potential}
\end{equation} 
where $j_1$ and $j_2$ are the first and last site in the wire, respectively.
This is illustrated in Fig.~\ref{fig:setup}.

The resulting linear conductance can be calculated directly from dynamical current-current correlation functions 
in the thermodynamic limit using Kubo's linear response theory~\cite{kubo57a}.
The dynamical DMRG method~\cite{jeck02a,jeck08b} can be used to compute the imaginary part of the frequency-resolved current-current correlation functions
for a finite lattice size $M$
\begin{equation}
G_{J,\eta}(\omega) = \left \langle 0 \left \vert J \frac{\eta}{(E_{0}-H+\hbar \omega)^2+\eta^2} J \right \vert 0 \right \rangle
\label{eq:correlation}
\end{equation}
where $H$ is the Hamilton operator, $\vert 0 \rangle$ is its ground state, $E_{0}$ is the ground-state energy, 
$J$ is the current operator in the wire, and $\eta > 0$ is a small number.
The Hamilton operator and the current operator depend on the system studied and thus will be given explicitly
in the next sections where we present our results for specific models.

It is important to realize, that the wire segment (and consequently both lead segments) is defined by the potential profile
(\ref{eq:potential}) and thus, in practice, by the lattice region, where the current operator $J$ acts.
It is possible but not necessary to include any information about the wire-lead layout in the Hamilton operator $H$,
such as different interactions or different degrees of freedom for wire and leads. 
In fact, in our first work we tested
our method with homogeneous systems only, i.e. the degrees of freedom and interactions were identical for all sites,
but for the presence of up to two impurities (on-site potentials) in the middle of the wire.

From the DMRG data for~(\ref{eq:correlation}) we can calculate the ``finite-size conductance''
\begin{equation}
G(M) = \left . \frac{q^2}{\omega} \left [ G_{J,\eta(M)}(\omega) - G_{J,\eta(M)}(-\omega) \right ] \right \vert_{\omega=0}
\label{eq:conductance}
\end{equation}
for a given system size $M$ using the scaling $\eta(M)=C/M$, where
$q$ is the charge carried by each fermion.
The constant $C$ depends on the properties of the system investigated.
For all results presented here we used $C=48t$ or $96t$.
Extrapolating $G(M)$ to infinite system length $M$ yields our estimation of the conductance $G$ in the thermodynamic limit.
Here, we used chain lengths up to $M=2000$.

For a noninteracting tight-binding chain it can be shown analytically that $G(M)$ converges to the exact result for $M\rightarrow\infty$,
i.e. the quantum of conductance 
\begin{equation}
G_0=n\frac{q^2}{h}
\label{eq:quantum}
\end{equation}
where $n=1$ for spinless fermions and $n=2$ for electrons.
For correlated systems we can only compute (\ref{eq:correlation}) and~(\ref{eq:conductance})
numerically.
In practice, there are two main practical issues. First, DMRG must be able to calculate 
$G(M)$ with enough accuracy for large system sizes $M$. Second, 
$G(M)$ should converge neatly to the actual value of the conductance for $M\gg 1$.
In our previous work we showed, that this method yields correct results for the renormalized
conductance of the one-component  Luttinger liquid in the homogeneous spinless fermion chain
as well as for one and two impurities added to the chain.
In the next sections we will show that this approach can also be applied to
inhomogeneous systems, electronic Hamiltonians and models with phonons.

In all our numerical results and figures we use $q=1$, and $\hbar=1$.
This yields $G_0 = \frac{n}{2\pi}$. Therefore, we show $2\pi G(M)$ in our figures.

\section{Inhomogeneous wire-lead system\label{sec:wire-lead}}

As mentioned above our method was demonstrated in Ref.~\cite{Bischoff2017}
for homogeneous systems only.
In real systems, however, wire and leads are of different nature. 
For instance, it is often assumed that metallic leads are Fermi liquids and thus they are modelled
by noninteracting tight-binding chains attached to the interacting
wire~\cite{meden03a,meden03b,Bohr2006,bohr07a,bran10,fabian10a,dias10,Mora16,Lang18}.
More generally, one may have to consider  
different interactions in the leads and in the wire as well as additional degrees of freedom (such as phonons)
in the wires.

Therefore, we have generalized our method to this situation and tested it on a spinless fermion model with different nearest-neighbor interactions
in the wire ($V_{\text{W}}$) and in the leads ($V_{\text{L}}$). Explicitly, this corresponds
to the Hamiltonian
\begin{align}
\label{eq:wire-lead}
H  = &  -t\sum\limits_{j=2}^{M}
\left ( c^{\dagger}_{j}c^{\phantom{\dagger}}_{j-1}+c^{\dagger}_{j-1}c^{\phantom{\dagger}}_{j} \right ) \\
&+ V_{\text{W}}\sum\limits_{j=j_{1}+1}^{j_{2}}\left ( n_{j} -\frac{1}{2} \right ) \left ( n_{j-1} -\frac{1}{2} \right )\nonumber\\
&+ V_{\text{L}}\sum\limits_{j=2}^{j_{1}}\left ( n_{j} -\frac{1}{2} \right ) \left ( n_{j-1} -\frac{1}{2} \right )\nonumber\\
&+ V_{\text{L}}\sum\limits_{j=j_{2+1}}^{M}\left ( n_{j} -\frac{1}{2} \right ) \left ( n_{j-1} -\frac{1}{2} \right ). \nonumber
\end{align}
The operator $c^{\dagger}_{j}$ ($c^{\phantom{\dagger}}_{j}$) creates (annihilates) a spinless fermion at position $j$ ($=1,\dots M$)
while $n_{j}=c^{\dagger}_{j}c^{\phantom{\dagger}}_{j}$ counts the number of spinless fermions on this site.
The hopping amplitude is identical for wire and leads and sets the energy unit $t=1$. 
We assume that the system has an even number of sites $M$ and is occupied by $M/2$ spinless fermions (half filling).

From the potential profile~(\ref{eq:potential}) follows that the current operator is given by       
\begin{equation}
J=\frac{1}{M_{W}-1} \frac{it}{\hbar}  \sum\limits_{j=j_{1}+1}^{j_{2}}
\left ( c^{\dagger}_{j}c^{\phantom{\dagger}}_{j-1}-c^{\dagger}_{j-1}c^{\phantom{\dagger}}_{j} \right ).\label{eq:current1}
\end{equation}
In principle, the wire edge sites $j_1$ and $j_2$ in the definition of the Hamiltonian~(\ref{eq:wire-lead})
could be different from those used for the potential profile (and thus the current operator).
Here we assume that the screening of the potential difference between source and drain occurs
entirely (and linearly) in the wire and thus we use the same values for $j_1$ and $j_2$ in both the Hamilton operator~(\ref{eq:wire-lead})
and the current operator~(\ref{eq:current1}).

\begin{figure}
\includegraphics[width=\columnwidth]{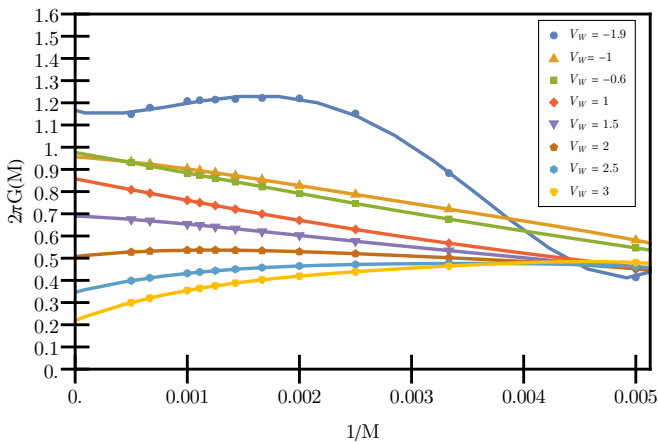}
\caption{
Scaling of the finite-system conductance $G(M)$ with the inverse system length $1/M$ for
spinless fermions in
a short interacting wire ($M_{\text{W}}=10$) between noninteracting leads ($V_{\text{L}}=0$).
The wire interaction strength is given in the inset.
Lines are polynomial fits.
}
\label{fig:noninteracting}
\end{figure}

Field-theoretical approaches predict that 
the conductance of a homogeneous Luttinger liquid is renormalized as 
\begin{equation}
G = K G_0
\label{eq:renormalization}
\end{equation}
where $K$ is the Luttinger parameter~\cite{apel82a,kane92a,Kane1992}.
We could reproduce this result in our previous work for the homogeneous spinless fermion model, 
i.e., for $V_{\text{W}}=V_{\text{L}}$ in~(\ref{eq:wire-lead}).
On the contrary, field-theoretical approaches predict that
the conductance through a Luttinger liquid wire connected 
to leads is not renormalized by interactions in the wire and is 
entirely determined by the lead properties~\cite{ponomarenko95a,safi95a,maslov95a,janz06,thomale11a}.
In particular, if the leads are Luttinger liquids their Luttinger parameter $K_{\text{L}}$
determines the conductance through~(\ref{eq:renormalization})~\cite{safi95a,janz06}.
However, field-theoretical approaches assume perfect contacts, i.e. without any single-particle
backscattering~\cite{janz06}. 
Numerical calculations for lattice models of quantum wires connected
to noninteracting leads show that the conductance deviates from the lead conductance $G_0$
when backscattering is taken into account.
In the lattice model~(\ref{eq:wire-lead}) 
single-particle backscattering is due to the sharp change of the on-site potential
at the boundaries between wire and leads
[$-V_{\text{W}}$ in the wire, $-V_{\text{L}}$ in the leads, and 
$-(V_{\text{W}}+V_{\text{L}})/2$ on the boundary sites $j_1$ and $j_2$],
which is required to maintain a constant density in equilibrium.

We have found that the calculation of the frequency-resolved correlation 
function~(\ref{eq:correlation}) with dynamical DMRG is not more difficult for an inhomogeneous system
($V_{\text{W}}\neq V_{\text{L}}$) than for a homogeneous one ($V_{\text{W}}= V_{\text{L}}$).
To investigate the conductance of a Luttinger liquid between leads, however, we 
would have to compute the conductance $G(M)$ in the limit of long wires $M_{\text{W}} \gg 1$
while maintaining much longer leads ($M \gg M_{\text{W}})$.
In our study of homogeneous systems~\cite{Bischoff2017} we found that the conductance scales
with $M_{\text{W}}/M$ and thus we were able to determine its value in the thermodynamic limit
using a fixed (and small) value of $M_{\text{W}}$. For the wire--lead system~(\ref{eq:wire-lead})
with $V_{\text{W}}\neq V_{\text{L}}$ considered here, we have found that the scaling 
with $M$ and $M_{\text{W}}$ (up to $M_{\text{W}}=82$) is much more complicated. As a consequence, it is not possible
to compute the conductance of an infinitely long wire in most cases. Thus we discuss
here the results obtained for short wires and present only results for $M_{\text{W}}=10$.

\begin{figure}
\includegraphics[width=.9\columnwidth]{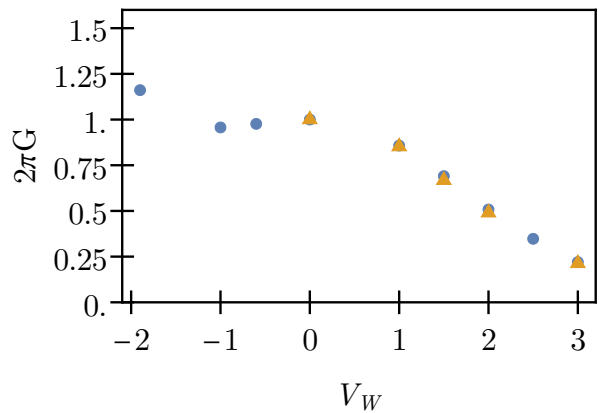}
\caption{
Conductance of a short interacting wire connected to noninteracting leads as a function of the
wire interaction $V_{\text{W}}$:
Extrapolated values $G = \lim_{M\rightarrow\infty} G(M)$ of the data in Fig.~\ref{fig:noninteracting}
for $M_{\text{W}}=10$ (circles) 
and DMRG data from Ref.~\cite{meden03a} for 12-site wires (triangles). 
}
\label{fig:schollwoeck}
\end{figure}

In Fig.~\ref{fig:noninteracting} we show the scaling of $G(M)$ with the system length $M$ for noninteracting
leads ($V_{\text{L}}=0$) and various interaction strengths $V_{\text{W}}$ in the wire.
The extrapolated values $G=\lim_{M\rightarrow \infty} G(M)$ are shown in Fig.~\ref{fig:schollwoeck}.
For repulsive interactions ($V_{\text{W}}> 0$) we observe a progressive decrease
of $G$ with increasing coupling $V_{\text{W}}$. 
The conductance of the spinless fermion model connected to noninteracting 
leads was investigated previously for repulsive interactions 
using DMRG and the functional renormalization group 
method~\cite{meden03a,meden03b}.
Our results for $G$ agree quantitatively with the (DMRG) results presented in Ref.~\cite{meden03a}
for a slightly longer wire ($M_{\text{W}}=12$), as shown in Fig.~\ref{fig:schollwoeck}.
In contrast, we observe in Fig.~\ref{fig:noninteracting}
that $G(M)$ tends to $G_0$ (i.e., the conductance of the noninteracting leads) 
for attractive interactions $0 > V_{\text{W}} > -2t$.
The non-monotonic behavior of $G(M)$ for $V_{\text{W}}=-1.9t$ in Fig.~\ref{fig:noninteracting} 
and the resulting large deviation of the extrapolated value from $G_0$ in Fig.~\ref{fig:schollwoeck} illustrate  
the problems that one encounters when analyzing finite-size effects in the wire-lead system. 

\begin{figure}
\includegraphics[width=.99\columnwidth]{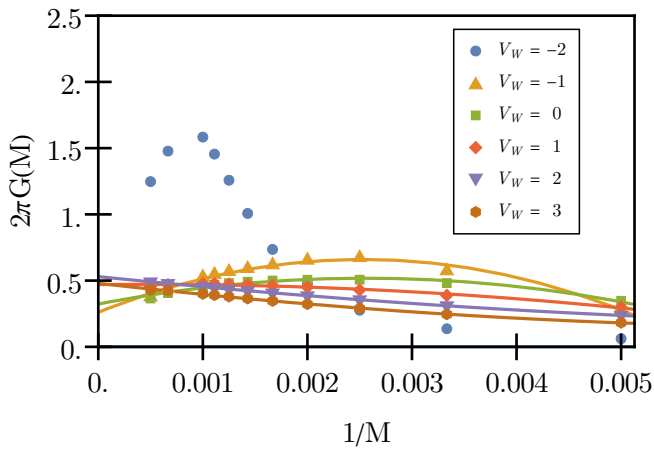}
\caption{
Same as Fig.~\ref{fig:noninteracting} but with a repulsive interaction $V_{\text{L}}=2t$ in the leads.  
}
\label{fig:repulsive}
\end{figure}

For an interaction $V_{\text{W}} > 2t$ the homogeneous half-filled spinless fermion model is in an insulating
charge-density-wave (CDW) phase and thus we expect that $G$ diminishes exponentially 
with increasing wire length $M_{\text{W}}$.
In the Luttinger liquid phase with repulsive interactions ($2t \geq V_{\text{W}} > 0$)
a power-law suppression of $G$ with increasing wire length was observed previously~\cite{meden03b}.
This deviation from the field-theoretical predictions mentioned above is due to the
single-particle backscattering caused by the sharp change of the interaction
at the wire boundaries.
The scaling of $G$ with the wire length in the attractive Luttinger liquid phase
($-2t < V_{\text{W}} < 0$) is not known.
As explained above, we can not determine the conductance for infinitely long wire
($M_{\text{W}} \rightarrow \infty$) with the present method.
Nevertheless, our data suggest that for attractive wire interactions within the Luttinger liquid regime the noninteracting leads 
determine the conductance of the system in agreement with field theory. This is in stark contrast to repulsive wire interactions, 
for which even a small $V_{\text{W}} > 0$ results in a reduction of $G$ from the noninteracting lead conductance $G_0$,
as already reported previously~\cite{janz06,meden03a,meden03b}.

While there are numerous studies of the conductance of wires connected to noninteracting 
leads~\cite{ponomarenko95a,safi95a,maslov95a,janz06,thomale11a,meden03a,meden03b,Bohr2006,bohr07a,bran10,fabian10a,dias10,Mora16,Lang18},
there seem to be only a few field-theoretical results for interacting leads~\cite{safi95a,janz06,Mora16}.
Figure~\ref{fig:repulsive} shows the scaling of $G(M)$ with the system size $M$ for leads with a repulsive
interaction $V_{\text{L}}=2t$. From the Bethe Ansatz solution of the homogeneous spinless fermion model 
we know that this corresponds to a Luttinger liquid parameter $K_{\text{L}}=1/2$~\cite{sirk12}
and thus to a conductance $G=G_0/2$ for perfect contacts according to field theory.
Clearly, all our results for $G(M)$ with $V_{\text{W}} > 0$ converge toward this value and thus agree
with the field-theoretical predictions.
Surprisingly, this also holds for the wire with $V_{\text{W}} = 3t$, which 
would correspond to a CDW insulator in the thermodynamic limit $M_{\text{W}} \rightarrow \infty$.
Therefore, we expect that $G$ will decrease with increasing $M_{\text{W}}$ 
for $V_{\text{W}} = 3t$.
For a noninteracting wire or an attractive interaction in the wire ($0 \geq V_{\text{W}} > -2t$), however,
we observe in Fig.~\ref{fig:repulsive}
a different, non-monotonic behavior of $G(M)$, which hinders the extrapolation of $G(M)$ to the limit $M \rightarrow \infty$.
Apparently, $G$ is lower than the lead conductance $G_0/2$ for $V_{\text{W}}=0$ and $-t$.

\begin{figure}
\includegraphics[width=.99\columnwidth]{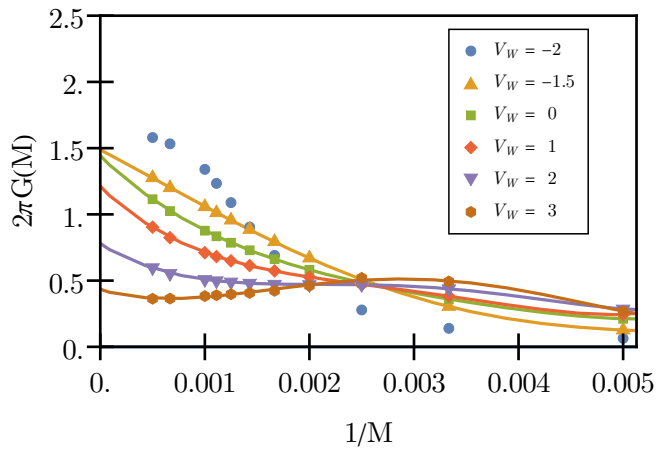}
\caption{
Same as Fig.~\ref{fig:noninteracting} but with an attractive interaction $V_{\text{L}}=-t$ in the leads.  
}
\label{fig:attractive}
\end{figure}

Figure~\ref{fig:attractive} shows the scaling of $G(M)$ with system length $M$ 
for leads with an attractive interaction $V_{\text{L}} = -t$.
This corresponds to a Luttinger liquid parameter $K_{\text{L}}=3/2$~\cite{sirk12}
and thus to a conductance $G=(3/2)G_0$ for perfect contacts according to field theory.
We see in Fig.~\ref{fig:attractive} that $G(M)$ seems to converge to different values for different 
repulsive couplings in the wire ($V_{\text{W}} > 0$).
Unfortunately, we cannot determine the limit $G=\lim_{M\rightarrow \infty} G(M)$
accurately even with system lengths up to $M=2000$.  
The values of $G(M)$ for $M=2000$ are above the conductance of an homogeneous
Luttinger liquid with the corresponding interaction $V_{\text{W}}$. They seem to increase further with $M$
but to converge to lower values than the lead conductance $G=(3/2)G_0$.
Therefore, it appears that the conductance of the wire-lead system is lower than the lead conductance in
that case, similarly to what is found for noninteracting leads due to single-particle backscattering.
For $V_{\text{W}} = 3t$ the conductance seems again to converge to a finite value. As for noninteracting leads 
it should vanish in the limit $M_{\text{W}}\rightarrow \infty$ as this interaction strength leads to 
a CDW insulating state in the wire. 
For a noninteracting wire and for attractive interactions in the wire ($0 \geq V_{\text{W}} \geq -2t$), however, we find 
that $G\approx (3/2)G_0$ as predicted by field theory but it is difficult to extrapolate $G(M)$ reliably 
close to the Luttinger phase boundary $V_{\text{W}} = -2t$.

In summary, our results agree with the known results for the conductance of the spinless
fermion wire-lead system and thus confirm the validity of our method.
Moreover, they also reveal complex qualitative differences in the attractive and repulsive
interaction regimes.
The conductance seems to be given by~(\ref{eq:renormalization}) with the lead Luttinger parameter
(as predicted by field theory) only when the interactions have the same sign in wire and leads.
The properties of short wires with attractive interactions connected to leads with repulsive
interactions (or vice versa) seem to be more complex than anticipated from the results reported so far
(using field theory) and should be investigated further.

\section{Holstein chain \label{sec:holstein}}

The interaction between charge carriers and phonons plays an important role for
transport properties~\cite{ziman60,mahan,solyom2}, in particular
for atomic or molecular wire junctions between leads~\cite{Agra02a,Agra02b,Ness02,Hiha12}. 
Therefore,
we have to generalize our method to models that include phonon degrees of freedom coupled
to the charge carriers. 
It is relatively straightforward to extend a DMRG program to models with phonons if one uses
a truncated eigenbasis of the boson number operator with at most $N_{\text{b}}$ states
to represent each phonon mode~\cite{jeck07}. The only real challenge is the computational cost that 
increases as $N^3_{\text{b}}$ and thus limit most applications to small $N_{\text{b}}$.
For Einstein phonons (i.e. dispersionless) and purely local couplings with the fermion degrees
of freedom 
it is possible to speed up DMRG calculations using a pseudo-site representation for bosons described in \cite{jeck98a}. 

\begin{figure}
\includegraphics[width=.99\columnwidth]{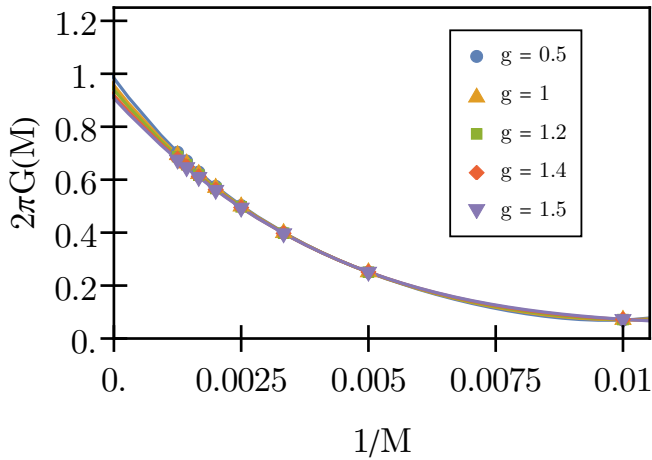}
\caption{
Scaling of the finite-system conductance $G(M)$ with the inverse system length $1/M$ for
the spinless fermion Holstein model in the adiabatic regime ($\omega_{\text{b}} = 0.1t$).
The coupling $g$ between fermions and phonons is given in the inset.
Lines are polynomial fits.
}
\label{fig:adiabatic}
\end{figure}

Therefore, we have tested our method on the simplest model with these properties, 
the spinless fermion Holstein model~\cite{hols59},
which is defined by the Hamiltonian
\begin{align}
H =&  -t\sum\limits_{j=2}^{M}
 \left ( c^{\dagger}_{j}c^{\phantom{\dagger}}_{j-1}+c^{\dagger}_{j-1}c^{\phantom{\dagger}}_{j} \right ) 
  + \omega_{\text{b}}\sum\limits_{j=1}^{M}b^{\dagger}_{j}b_{j}\nonumber\\
  &-g \, \omega_{\text{b}}\sum\limits_{j=1}^{M}\left(b^{\dagger}_{j}+b_{j}\right)n_{j},\label{eq:HolsteinHam}
\end{align}
where $\omega_{\text{b}}$ is the frequency of the Einstein phonons created (annihilated) by the boson operators $b^{\dagger}_{j}$ ($b_{j}$) 
and $g$ is the dimensionless coupling between spinless fermions and phonons. 
The spinless fermion operators $c^{\dagger}_{j}$, $c_{j}$, and $n_{j}$ have the same meaning as in the previous section
and the hopping term again sets the energy unit $t=1$, while we consider a system with an even number of sites $M$ filled
with $M/2$ spinless fermions.
This Hamiltonian is homogeneous and thus the wire and lead sections are defined by the potential profile~(\ref{eq:potential}) only.
The current operator is again given by~(\ref{eq:current1}).

\begin{figure}
\includegraphics[width=.99\columnwidth]{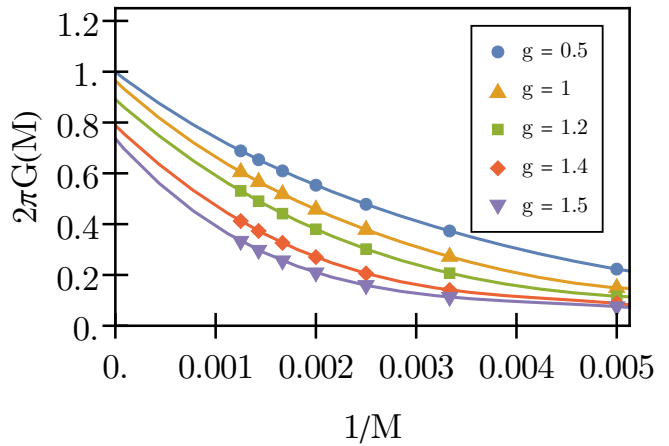}
\caption{
Same as Fig.~\ref{fig:adiabatic} but for an intermediate phonon frequency $\omega_{\text{b}}=t$. 
}
\label{fig:nonadiabatic}
\end{figure}

We have found that calculating the conductance with our method
is more difficult for the Holstein model than for purely fermionic models such as~(\ref{eq:wire-lead}).
This is due not only to the increase of the computational cost with the phonon cutoff $N_{\text{b}}$
but also to the fact that the dynamical DMRG algorithm~\cite{jeck02a,jeck08b} often fails to converge in a reasonable time 
for large $M$. Thus we can compute $G(M)$ systematically for smaller system sizes $M$ than in fermionic systems
and we show here results for $M$ up to $800$ sites only.
(One calculation for this system size requires about 600 CPU hours and 4GB of memory. Thus
one could certainly treat larger systems using supercomputer facilities.)
In contrast, the scaling of $G(M)$ 
with the system size is regular in the homogeneous Holstein wire as found previously for homogeneous
fermionic systems~\cite{Bischoff2017}. Thus we can estimate the conductance $G$ in the thermodynamic limit despite the short system lengths
as illustrated in Fig.~\ref{fig:adiabatic} for the adiabatic regime ($\omega_{\text{b}} = 0.1t$)
and in Fig.~\ref{fig:nonadiabatic} 
for the intermediate regime ($\omega_{\text{b}} = t$).

The phase diagram of the half-filled spinless fermion Holstein model 
was determined previously with DMRG~\cite{ejim09} and quantum Monte Carlo computations~\cite{greit15}.
It exhibits a Luttinger liquid phase at weak coupling or high phonon frequency.
The Luttinger liquid parameter $K$ was calculated from the long-wavelength limit
of the static structure factor~\cite{ejim09,greit15}. This corresponds to the
parameter $K$ determining the exponents in the power-law correlation functions
of the Luttinger liquid theory for purely fermionic systems~\cite{giamarchi03}.

\begin{figure}
\includegraphics[width=.9\columnwidth]{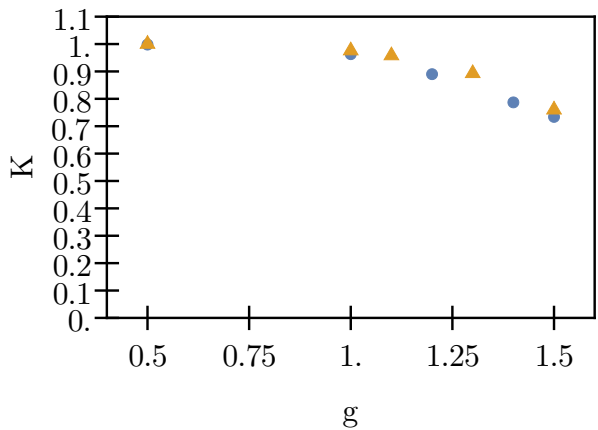}
\caption{
Luttinger parameter $K$ of the spinless fermion Holstein model with $\omega=t$ as a function of the coupling $g$.
$K$ was determined from the extrapolated value of the conductance $G$ in Fig.~\ref{fig:nonadiabatic} (circles)
and from the charge structure factor in Ref.~\cite{ejim09} (triangles).
}
\label{fig:ejima}
\end{figure}

As seen in Fig.~\ref{fig:adiabatic}, 
we have found that the conductance corresponds to $K=G/G_0 \approx 1$ in the adiabatic regime $\omega_{\text{b}} = 0.1t$
for a coupling up to at least $g=1.5$ in agreement with the values determined from the
structure factor~\cite{ejim09,greit15}.
In the intermediate regime $\omega_{\text{b}} = t$ shown in Fig.~\ref{fig:nonadiabatic}, the conductance $G(M)$
clearly decreases with increasing coupling $g$. The extrapolated values of $G$ for $1/M \rightarrow 0$
yield parameters $K=G/G_0$ that agree well with the values determined from
the structure factor in Ref.~\cite{ejim09}, as shown in Fig.~\ref{fig:ejima}.
In particular, we do not observe any sign of superconducting fluctuations ($K > 1$ or $G>G_0$). 
Unfortunately, we are not able to determine the conductance close
to the metal-insulator transition because this requires too much computational resources 
(the phonon cutoff $N_{\text{b}}$ should be larger than the value $N_{\text{b}}=4$ used here).

In summary, our results for the conductance in the spinless fermion Holstein model
agree with the Luttinger parameter $K$ defined from the power-law decay of correlation functions
(or equivalently from the static structure factor).
This confirms the validity of our method for models with charge carriers
coupled to phonon degrees of freedom. It also shows that the relation~(\ref{eq:renormalization})
for the conductance,
which was derived for purely fermionic systems using field theory~\cite{giamarchi03,apel82a,kane92a,Kane1992},
remains valid for the Luttinger liquid phase of models including phonons.

\section{Hubbard chain \label{sec:hubbard}}

A necessary but simple extension of our method is the generalization to electrons (i.e., fermions with spin $\frac{1}{2}$).
To illustrate this extension we consider the one-dimensional homogeneous Hubbard chain
with the Hamiltonian
\begin{equation}
H =  -t\sum\limits_{j=2}^{M} \sum_{\sigma}
 \left ( c^{\dagger}_{j,\sigma}c^{\phantom{\dagger}}_{j-1,\sigma}+
 \text{h.c.}
\right ) 
+ U\sum\limits_{j=1}^{M}n_{j,\uparrow}n_{j,\downarrow}\label{eq:HubbardHam},
\end{equation}
where $c^{\dagger}_{j,\sigma}$ ($c^{\phantom{\dagger}}_{j,\sigma}$) creates (annihilates) an electron with spin $\sigma (= \uparrow,\downarrow)$
on site $j$ and $n_{j,\sigma}=c^{\dagger}_{j,\sigma} c^{\phantom{\dagger}}_{j,\sigma}$ is the corresponding particle number operator.
The strength of the on-site coupling between electrons is given by the Hubbard parameter $U$
and the hopping term again sets the energy unit $t=1$.
We assume that the system length $M$ is even and that the system contains $M/2$ electrons of each spin (half filling).

The current operator for the electrons with spin $\sigma$ in the wire is
a simple generalization of~(\ref{eq:current1})
\begin{equation}
J_{\sigma}=\frac{1}{M_{\text{W}}-1} \frac{it}{\hbar}  \sum\limits_{j=j_{1}+1}^{j_{2}} 
\left ( c^{\dagger}_{j,\sigma}c^{\phantom{\dagger}}_{j-1,\sigma}-c^{\dagger}_{j-1,\sigma}c^{\phantom{\dagger}}_{j,\sigma} \right ) .
\label{eq:current2}
\end{equation}
The current operator for the charge transport is then given by 
\begin{equation}
J_{\text{C}} = J_{\uparrow} + J_{\downarrow}.
\end{equation}
Using this definition in Eqs.~(\ref{eq:correlation}) and~(\ref{eq:conductance})
we obtain the (charge) conductance $G(M)$ of the electronic system.
The current operator for the spin transport is similarly given by 
\begin{equation}
J_{\text{S}} = J_{\uparrow} - J_{\downarrow}.
\end{equation}
Using this definition in Eqs.~(\ref{eq:correlation}) and~(\ref{eq:conductance})
we obtain the spin conductance $G_{\text{S}}(M)$ of the electronic system.
This describes the linear response of the system to an external magnetic field with 
the profile~(\ref{eq:potential}).

The one-dimensional Hubbard model is exactly solvable using the Bethe Ansatz method~\cite{Lieb68,essler05}.
Its low-energy properties are characterized by a separation of charge and spin excitations
and its gapless excitation modes are described by the Luttinger liquid theory.
At half filling there is an exact symmetry between the charge (spin) properties for $U$ and
the spin (charge) properties for a coupling $-U$. In particular, the ground state is a Mott insulator
with gapped charge excitations and gapless spin excitations for $U>0$, while 
it is a Luther-Emery liquid~\cite{giamarchi03} with gapless charge excitations and gapped spin excitations for $U < 0$.

We have found that calculating the (charge or spin) conductance with our method is only slightly more difficult 
for the homogeneous electronic model~(\ref{eq:HubbardHam}) with $U \neq 0$ than for the homogeneous spinless fermion model.
This is understandable because in both cases there is only one gapless excitation mode in the system.
As expected the charge conductance $G(M)$ for any $U$ is equal within the numerical errors
to the spin conductance  $G_{\text{S}}(M)$ for an interaction $-U$. Thus we discuss only the latter case.

\begin{figure}
\includegraphics[width=.99\columnwidth]{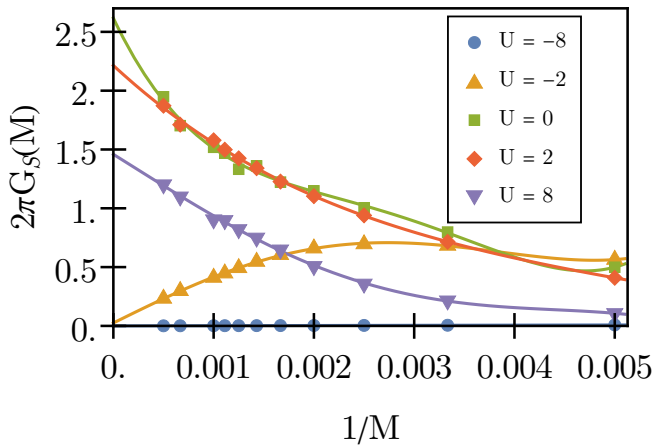}
\caption{
Spin conductance $G_{\text{S}}(M)$ of the homogeneous Hubbard chain as a function of the inverse system length $1/M$.
The Hubbard interaction $U$ is given in the legend.
Lines are polynomial fits.
}
\label{fig:spin}
\end{figure}

In Fig.~\ref{fig:spin} the scaling of $G_{\text{S}}(M)$ with $M$  is plotted for several values of the interaction $U$. 
For noninteracting electrons $G_{\text{S}}(M)$ approaches $G_0$ as expected. [Note that the quantum of conductance~(\ref{eq:quantum}) 
for electrons ($n=2$) equals $2 \pi G_0 = 2$ in our units].
The scattering of the $G_{\text{S}}(M)$ data and the resulting poor extrapolation are due to the lesser
accuracy of DMRG for a chain with two gapless excitation modes (i.e., for $U=0$) than for a single one (i.e. for $U\neq 0$).
For $U<0$, $G_{\text{S}}(M)$ vanishes in the thermodynamic limit as required for a phase with gapped spin excitations.
The qualitatively different behaviors for weak and strong attractive interactions is probably due to 
the different correlation length, which is smaller than $M_{\text{W}}=10$ for $U=-8t$ 
but larger for $U=-2t$. 
The Luttinger liquid parameter for the spinless spin excitations  is $K_{\text{S}}=1$ in the Mott insulating phase at half 
filling for any $U>0$.
Concordantly, we observe in Fig.~\ref{fig:spin} that $G_{\text{S}}(M)$ approaches $G_0$ for a weak repulsive interaction ($U=2t$).
For stronger interactions the convergence of $G_{\text{S}}(M)$ toward $G_0$ is less clear as shown by the results
for $U=8t$ in Fig.~\ref{fig:spin}. 
This is due to the small band width of the gapless spin excitations (of the order of $4t^2/U$ for large $U$)
which requires a smaller broadening $\eta \sim 4t^2/U$ and thus a larger $M$ to reach the same resolution
for the dynamical correlation functions~(\ref{eq:correlation}).

\section{Conclusion and Outlook \label{sec:conclusion}}

We have extended the DMRG method for the linear conductance of one-dimensional correlated lattice models to 
more complex systems including interacting leads, coupling to phonons, and the electron spin properties.
The tests conducted in this work reveal 
intriguing differences
when the sign of the interaction (attractive/repulsive) differs in wire and leads
and confirm that the renormalization of the conductance by the Luttinger parameter~(\ref{eq:renormalization})
remains valid for the Luttinger liquid phase of homogeneous models with coupling to phonons.
Unfortunately, we have found that we cannot determine the conductance of a
Luttinger liquid wire between leads with a different interaction but can only investigate short
wires because of the complicated finite-size scaling.
Moreover, for models with coupling to phonon degrees of freedom the high computational cost of
dynamical DMRG calculations limits the system sizes that can be used. 

Nevertheless,
the extensions presented in this work will certainly allows us to compute the conductance
of short electron-phonon-coupled wires connected to interacting leads (without phonons).
Therefore, we will be able to study more realistic models for realizations of quasi-one-dimensional 
electronic conductors such as  
atomic or molecular wire junctions~\cite{Agra02a,Agra02b,Ness02,Hiha12}. 
Moreover, we point out that our approach is not limited to the charge and spin conductance but could be extended to other 
transport properties that are described by a local current operator, such as the energy current~\cite{zotos97,heid02,stei16,karrasch16a}.

\begin{acknowledgments}
Jan Bischoff would like to thank the Lower Saxony PhD-Programme \textit{Contacts in Nanosystems} for financial support.
We also acknowledge support from the DFG (Deutsche Forschungsgemeinschaft) through Grant No. JE 261/2-2 in the
Research Unit \textit{Advanced Computational Methods for Strongly Correlated Quantum Systems} (FOR 1807).
The cluster system at the Leibniz Universit\"{a}t Hannover was used for the computations.
\end{acknowledgments}

\bibliographystyle{biblev1}
\bibliography{mybibliography}{}

\end{document}